\documentclass[twocolumn]{article}
\usepackage[utf8]{inputenc}
\usepackage[affil-it]{authblk}
\usepackage{pgfplots}
\usepackage{subfig}
\usepackage{mathtools}
\usepackage{graphicx,float}
\usepackage{siunitx}
\DeclareSIUnit\inch{in}
\DeclareSIUnit\rpm{RPM}
\DeclareSIUnit\year{yr}
\usepackage{amsmath}
\usepackage{amsfonts}
\usepackage{amssymb}

\usepackage{hyperref}
\hypersetup{
    colorlinks=true,
    linkcolor=blue,
    filecolor=magenta,      
    urlcolor=cyan,
}

\pgfplotsset{width=8cm,compat=1.9}
\usepackage{braket}
\numberwithin{equation}{section}

\title{\textbf{Dynamics and Statistical Mechanics of Closed Composite Multi-Level University and Scientific Hierarchical Systems Pertaining to the Undergraduate and Graduate Student Phenomena}}

\author{Jason Garver}
\affil{University Of Minnesota}
\date{\today}
\begin{document}

\twocolumn[
\begin{@twocolumnfalse}
\maketitle
\begin{abstract}
Content: Qualitative and ad-hoc descriptions of observations of the ``undergraduate'' and ``graduate'' effects. Comparisons between the classical models of both undergraduate and graduate systems separately lead to the conclusion that a quantized model is inevitably required to explain certain effects such as ``Spooky Action At A Seminar'' and the FTR (Food Travel Ratio) among other select observed phenomena. Two regimes of dynamics in systems are explored as well as interactions of composite systems. An attempt is made to refine the undergraduate/graduate statistical theory in order to explain the common equilibrium systems known as ``labs''. Finally this work will begin the discussion on the mysterious ultra-long range ``Advisor'' interaction, which seems to contradict the theory of undergraduate/graduate interaction as well as specific graduate/graduate interactions, and graduate/research interactions in erratic ways.
\end{abstract}
\end{@twocolumnfalse}
]

\section{Introduction}
Since the founding of the University of Bologna in 1088, a series of observations have been made which have lead to a theory of the dynamics and interactions of select isolated systems which came to be known as ``University''. It is of interest to study the constituents of these systems in detail to develop the conditions for equilibrium, and an equation of state that will accurately describe how such systems evolve in time. Furthermore, there are well documented observations that we shall try to understand with the principles of studentary kinematics and dynamics. We will however leave this classical regime to attempt to make sense of several observable events that cannot be explained with a completely classical theory.

\subsection{ Definitions }
In this work we shall study only closed systems, which carry a fixed number of constituents at any one time. However, some mention of open systems will be necessary for a variety of reasons that I will postpone until needed. Let us first become acquainted with what is under study before giving ourselves more advanced descriptions. 

\begin{enumerate}
    \item University: The ``Univrersity'' is the generic term for our system in question. These systems need not always be the same size, as all work here is easily generalized\footnote{This generalization fails under differing regimes, to be discussed later with open systems.}. 
    
    \item Undergraduate/graduate (UG/GS's): The basic quanta of the university system, we will see through failures of classical theory that the basic macroscopic particle ``student'' is further composed of UG/GS's
\end{enumerate}

The majority of this work will deal with the university system and it's constituents, with all other interactions/constituents defined when needed. Note, the university systems discussed here are idealized and realistically will require further study of their constituent's behavior for completely accurate predictions. \footnote{The fields of Biology and Psychology currently explore these parameters}

\section{Student Dynamics}
To begin our discussion on the dynamics of students (that is, subsystems of UG's or G's), I want to introduce several pertinent parameters then present the three universal (or so you might think!) Laws of Education from which all motion may be described. 

\subsection{Parameters}
We must initially define two parameters ad-hoc, but we will soon see they are quite natural aspects of the theory. Firstly, we define the ``progress'' $r$\footnote{Progress is historically denoted $r$ to avoid any associations with the letter $P$ like ``Pleasure'' or ``Price''}. Progress is quite simply the total lengths a student must go in the educational regime minus how far they've already gone. Derivatives of progress are as follows.

\begin{align}
   \frac{\mathrm{d}r}{\mathrm{d}t} = v \qquad \frac{\mathrm{d}v}{\mathrm{d}t} = a
\end{align}

Where $v$ is the rate in which a student progresses\footnote{Sometimes termed ``Productivity''}, and $a$ is the student's attitude.
\\

Lastly the motivation $m$ is defined as a student's willingness to change their rate of progression with its reciprocal the ``Procrastination'' $\mu$ being the resistance to progress.

\subsection{The Laws of Educational Motion}
We are now in a position to empirically state The Laws of (student) Educational Motion.

\begin{enumerate}
    \item A student who is in a state of progress will remain in that state of progress unless acted upon by an outside influence, and likewise a sleeping student will remain so by the same conditions.
    
    \item The rate in which a student's tendency to make progress changes is proportional to the student's attitude.
    
    \item If an outside influence is applied, a student will always experience an equal and opposite (negative) influence.
\end{enumerate}

These laws are affectionately abbreviated ``eN1L'', ``eN2L'', ``eN3L'' because of their similarity to Newton's Laws of Motion. Luckily the similarity does not end, as we may apply the same machinery from Newtonian mechanics to Educational Mechanics.
\\

\subsection{Analysis of The Laws of Educational Motion}
To begin, let us define the student's ``momentum'' \(p\) as the rate in which a student will make progress scaled by their motivation (or procrastination).

\begin{align}
    p &= m \dot{r}
\end{align}

Which gives us an expression for eN2L recognizing that $m$ was the constant of proportionality mentioned before.

\begin{align}
    \frac{\mathrm{d} p }{\mathrm{d}t} &= m a \label{eq:2}
\end{align}

Equation \eqref{eq:2} tells us interesting behavior of students. We see that a student with high motivation (low procrastination) and high attitude will have a higher rate of change in their progress. It will be convenient now to define the ``force'' \(f\)\footnote{It is not a coincidence that the motivation dependence on force seems to be the same regardless of interaction. As it turns out ``intrinsic'' and ``extrinsic'' motivations blur in the University system but the theory does not require that behavior} per student.

\begin{align}
    \frac{\mathrm{d} p}{\mathrm{d}t} &= m a = f\label{eq5}
\end{align}

Now we are in a position to analyze eN1L, the outside influence mentioned is the force we just defined in \eqref{eq5}, thus students only change how fast they make progress when they're forced to.
\\

Similarly, eN3L states that whenever a student is acted on by a force, there will be an equal and opposite force produced on the student (opposite in that the opposite effect on the progress is produced). We will discuss these forces later, but for now let us derive the equations of (educational) motion for students.

\subsection{Equations of Motion}
By taking the appropriate integrals (and remembering that $m$ and $a$ may be both complicated functions of time we can produce (in parallel to Newtonian Mechanics)) equations educational motion. However, it should be noted that these equations describe motion in the educational regime only, but are however not disconnected from Newtonian Motion.
\\

It is for this reason that we seek to combine Newtonian Mechanics with Educational Mechanics, which is easily done in the following way. Consider all of the usual Newtonian parameters, then all of the Educational parameters discussed (designated by a subscript of ``$e$'') we can produce by inspection the following.

\begin{align}
\begin{split}
    \vec{r}_{\text{tot}} &= \vec{r} + r_{e}
    \\
    \vec{p}_{\text{tot}} &= \vec{p} + p_{e}
    \\
    \vec{F}_{\text{tot}} &=  \vec{f}_{\text{net}} +  f_{e}
    \end{split}\label{eq2.6}
\end{align}
With all other related quantities left to the reader.

\subsection{Dimensions}
Until this point I have not mentioned the units or dimensions of the quantities discussed. Like the ``Radian'', the education parameters in equation \eqref{eq2.6} have no dimensions. However it is sometime useful to define $[ r_{e} ] = \text{``Creditmeter''}$, with the derivatives having the appropriate \SI{}{creditmeter \per \second}, \SI{}{creditmeter \per \second \squared}.

\section{Interaction in closed - Isolated University Systems}
We now direct our discussion to the motion and interactions of UG's and Gs's within the closed University system.

\subsection{Undergraduate Observations and Dynamics}
Firstly we will start with the UG motion and UG/UG interactions. UGs are subject two only a handful of main forces; ``Midterms'', ``Graduation'', and the umbrella termed ``Collegiate Experience/Fun'' (CEF).

\subsubsection*{Midterm and Graduation Forces}
The UG's main changes in momentum are from the midterm and graduation forces. Both of these interactions are infinite in range, and unlike many other long-range forces they do not decay over distance, but have temporal dependence. The forms of these forces are as follows.

\begin{align}
f_{\text{mt}} =  \mu \sin(2\pi\, \nu\, t) \quad f_{g} = s e^{-s t} 
\end{align}

Where $\nu$ is the frequency of the ``midterm'' effect, and $s$ is the ``Cocky Stress Index''. As we can see, the midterm force is periodic and has magnitude scaled by the UG's procrastination. The graduation force is likewise uninteresting, the student's stress index ($ s \in (0, \infty)$) ultimately determines the rate of attitude decay.

\subsubsection*{CEF Force}
To develop a qualitative description of the CEF force, consider one UG in a university system, this UG will not be subject to CEF forces, so according to eN1L, it will stay in the same state of progress. Now imagine you inject several hundred UGs into the system, how does the first UG respond? 
\\

The magnitude of the CEF force is proportional to the number density of UGs, however it's range is limited to the distance between the UG and special geometric features called ``Student Unions'' and ``Dorms''. Outside of these distances, the CEF force decays wildly and is un-predictable. Classically, the CEF cannot be explained in detail, nor does it have a functional form.
\\

Given these force interactions we can summarize UG motion classically in the following way. UG's are always acted on by the midterm and graduation forces and thus have simple solutions to their equations of educational motion\footnote{Ignoring UG/UG interactions}. The CEF force is not completely defined classically, but we know there is spatial dependence due to student unions and dorms that will be discussed in the Quantum Theory of Students.

\subsection{Graduate Dynamics}
In contrast to UG's, GS's are much more difficult to study. In general, a GS has much lower motivation and attitude, and are thus difficult to detect freely in the University. Experimental results conclude that GS have two semi-classical driving forces; ``Free Food'' and ''Thesis''\footnote{in higher motivation regimes this is called ``Research''}. These effects are poorly understood using classical theory, and thus are postponed until a non-classical theory is presented. Lastly, GS's are subject to a mysterious long-range, ``Advisor'' force, which tends to dramatically lower attitude and increase motivation exponentially over some characteristic timescale.

\section{The Theory of Relative Education}
Though not the focus of this paper, it is prudent to develop several more classical ideas and explore the relative nature of education.
\\

\subsection{Energy, Work}
To preface the discussion on relative education, we will be needing several other important quantities.
\\

The progression energy of a student is defined in the following way.

\begin{align}
    T_{e} = \frac{p_{e}^2}{2 m_{e}} \Leftrightarrow \frac{p_{e}^{2}\mu}{2}\label{eq4.1}
\end{align}
Equation \eqref{eq4.1} shows that a student's energy is directly related to their motivation, hence a student with low motivation will likely have low energy, etc. Students of both kinds expend energy in processes called ``Work''\footnote{In the UG case this is called ``Study'', in the GS case it's called ``Research'' (which can be confusing)}. The work is defined as the net force acting on a student multiplied by an infinitesimal amount of progress. 

\begin{align}
    \mathrm{d}W_{e}^{d} &=f_{e} \mathrm{d}r \label{eq4.2}
\end{align}
Where I have designated work with a superscript $d$ because work a student does is never conservative which is known as the ``Law of No Purpose''. Work happens to be related to the total change in a student's energy in the following way

\begin{align}
    \Delta E_{e} &= -W_{e}
\end{align}

Notice the placement of the minus sign, this is because as a student does more work, their energy decreases.\footnote{Negative work is commonly referred to as ``Relaxation''}
In the end a student has total energy (which is by no means conserved),
\begin{align}
    T_{e} + V_{e} = E_{e}
\end{align}

where $V$ is a student's potential\footnote{Common in the popular media ``This student had so much potential before University...''}. 

\subsection{The Special Theory of Education}
In 1905 it was realized by Einstein that a student's perception of time and space are not absolute, and instead change as the student progresses faster or slower. For example Imagine a student sitting on a train while you observe the train from another train. From the train in which you reside, the student seems to be studying for some time $t_{\text{train}}$ (the length of the train ride). The student however, finds that they've been on the train (studying) for a time $t_{p} < t_{\text{train}}$. This is because a student who has a higher rate of progress (than you, sitting in the other train not studying) will experience a higher rate of the passage of time (likewise a student that does not make progress very quickly will find time dragging on and on). This effect can be mathematically presented with the ``Lorentz Gamma-e Factor''.

\begin{align}
    \gamma_{e} &= \frac{1}{\sqrt{1-(\dot{r_{e}}/c)^2}}
\end{align}

Where $c$ is a constant usually called ``coffee'' or ``concentration constant'' present in all relativistic student systems.\footnote{The ratio $\dot{r_{e}}/c = 1$ when a student is progressing as fast as one composed entirely of ``coffee''}

Thus the time interval as experienced by the student studying $t'$ and the non-studying student's interval $t$ is

\begin{align}
    t' &= \frac{t}{\gamma_{e}}
\end{align}

In the limiting cases, we see if a student is not studying, the passage of time is the same for all students. If a student is studying to the ``$c$'' limit, we see that they experience little no passage of time (rather, all sense of time is ignored).
\\

Similarly, students who are studying perceive spatial intervals differently. Since a studying student has trouble studying over distance, a high rate of progress produces the effect that distances are dragged out and extended\footnote{Commonly heard from GS's ``The coffee shop is SO far away''} thus we produce a similar relationship.

\begin{align}
    d' &= \gamma d
\end{align}
Finally, in this regime, using the last two relations and setting $c=1$, a student's energy becomes the following (proof is left to the reader).

\begin{align}
    E_{e} &= p_{e} + m_{0}
\end{align}

where $m_{0}$ is the motivation of a student who is not studying, and  the relativistic momentum is $p_{e} = (\dot{r} m_{0})/ \gamma$.

\subsection{The General Theory of Education}
Some time after the Special Theory was published (Einstein was hard at work, so nobody can hope to measure this time interval) a beautiful geometric description of the progress of students was found. Einstein realized that students who procrastinate heavily, tend to change the progress of other nearby students, causing them to also procrastinate heavily.
\\

To understand this, imagine a student who procrastinates much more than other students around them. This student creates a deformation in the ``P-field'' (the progress-time continuum), which draws in other students, and those students further deform the field drawing in more and more students, etc, etc. Some students make progress so rapidly that they only deflect around the strong deformations in the P-field. The mathematics of these effects are beyond the scope of this paper (which involve several dozen coupled equations, that would require the coffee-limit rate of progress to derive).
\\

One consequence of this effect is the CEF force. It was mentioned that UG's often experience this CEF force in locations full of other UG's. We now see that the CEF is simply the potential wells created by deformations of the P-field. Students in these large (negative) potentials tend to lose motivation and thus progress is slowed or stopped. In the GS case, the UG's are not the source, and instead other GS's and/or a mysterious effect dubbed ``Advisor Vacation''. These potential wells for GS's effect the GS motivation much more and thus are called ``Procrastination Holes''. A GS trapped within a Procrastination Hole rarely escapes to do any work, as they just don't have the motivation.

\section{Quantum Theory of Students and Fields}
Having drawn as much from the classical theory of students as we can, we now shift our gaze to the weird and mysterious quantum theory of UG's and GS's. As usual, we want to define the limits of the regime we're working in, and as such it should be noted that the quantum theory applies only to low motivation students who have also made considerable progress\footnote{This means GS's are much more accurately described by quantum theories than UG's}. The goals of this theory will be to accurately describe the state in which a student exists given a set of parameters, and identify possible solutions to unsolved problems such as UG/UG interaction, GS/GS, and GS/UG interaction, the Advisor interaction, ``Spooky Action at a Seminar'', and the FTR (Food travel ratio).

\subsection{The Wavefunction and Postulate of Quantum Student Theory}
The most basic ideas of QST (quantum student theory) are that a student's uncertainty of momentum and progress cannot be both known beyond a certain fundamental limit

\begin{align}
    \sigma_{p}\sigma_{r} \geq \frac{\hbar}{2}
\end{align}

As such, if you know how far along a student is, you don't know how fast they're progressing, etc. Using this idea, the Schr\"{o}dinger's equation for the ''wavefunction'' ($\psi$, pronounced ``sigh'') which describes all physical quantities of the student quanta is formulated. 

\begin{align}
    i \hbar \frac{\partial}{\partial t} \psi(r,t) &=  V_{e}(r,t)\psi(r,t)-\frac{\hbar^{2}}{2m_{e}}\nabla^{2}\psi(r,t) \label{eq5.2}
\end{align}
Where $\hbar$ is Planck's constant for reducing students.

\subsection{UG/UG interactions}
Solving equation \eqref{eq5.2} can be a daunting experience, so we will qualitatively speak about the UG/UG interaction. Firstly it should be noted that the UG is semi-unbound spatially. Experimental results show that while UGs do tend to ``clump'' into groups of 50 or more, these clumpings only last for a finite time. The density of students in one of these clumps as a function of time is as follows.

\begin{align}
    \rho_{ug} &= \frac{(t^{2} / \tau^{2})}{e^{t^{2}/\tau^{2}} -1} \delta
\end{align}
Where $\delta$ is the critical density of UGs in the clump. This effect has been dubbed ``class'', and usually has a characteristic timescale $\tau= \SI{50}{\minute}$, though longer times have been observed.
\\

There is another UG effect with strong temporal dependence that must be discussed, which is UG decay. It has been discovered that UGs decay into GSs\footnote{Some UG decay actually removes them from the system, which hints to non-closed nature of University} after about $\tau_{\text{decay}} \approx \SI{4}{\year}$. Shortly prior to decay, UGs exhibit higher than normal motivation, but directly after decay motivation drops asymptotically to zero.
\\

UG/UG interactions are extremely common though the effect on motivation and momentum is not considerable compared to CEF or midterm effects. Every UG produces invisible lines of influence called the ``P-M Field''. Imagine every UG has a charge $\mathcal{Q}$, a positive $\mathcal{Q}$ produces an m-field, that tends to motivate other nearby UGs, and a negative $\mathcal{Q}$ tends to cause other UGs to become distracted. Now imagine that a UG interacts via the P-M interaction with another UG, both have $+\mathcal{Q}$, these two UG seem to increase their total motivation, and momentum while the interaction continues. Experts have dubbed this ``group studying'' - it is observed that groups of up to ten UGs may benefit from this interaction, but larger systems tend to result in diminishing returns. Other combinations of $\mathcal{Q}$ have similar results, and are left as an exercise to the reader. 
\\

Lastly, UGs experience a scattering phenomenon. While unbound, and uncharged ($\mathcal{Q}_{\text{tot}} \sim 0$), certain combinations of spatial location, total progress, and the Cocky Stress index $s$ discussed earlier allow UGs to bind in small pairs, or triplets for short times before scattering off each other. This effect is not fully understood, as the total momentum and motivation of the UGs are conserved, and there does not seem to result in a change in rate of progress due to this interaction (and thus is not-not, nor is it energetically favorable!). Experiments do show that the interaction is more likely to occur before and after a class interaction.

\subsection{GS/GS Interaction}
The GS/GS interaction is far, far more complicated than the UG cases, let's first look at any similarities we can produce between them. First and foremost, the GS does experience a class interaction like the UG, however the form must be modified.

\begin{align}
    \rho_{gs} &= \frac{(t^{2} / \tau^{2})}{e^{t^{2}/\tau^{2}} -1} \delta H(2-t_{p})
\end{align}

Where $H$ is the Heaviside function, $t_{p}$ is the total length of time a GS has existed in the system (since UG decay), and the characteristic time $\tau = \SI{90}{\minute}$. It seems that after a finite time ($t_{p} \approx \SI{2}{\year}$), GS cease respond to the class interaction.
\\

Next we would like to know if the GS decays like the UG - the answer is undoubtedly yes, but what is the decay time? And what is the decay product? These questions are largely speculative, like the UG, some GS decay ``out'' of the university system entirely. Some GS decay into a slightly more useful form that we'll call ``Post-Doc'', and yet it seems that some GS do not decay whatsoever, and are stable on even cosmic timescales. We lack an exact functional form for this decay, but we know it is on timescales $t_{\text{Ph.d}} \sim \SI{8}{\year}$.
\\

GS, unlike the UG are very strongly bound spatially to areas of low motivation called ``labs''. Starting from the time a UG decays into a GS, a strong interaction called ``research'' takes over (mostly replacing CEF). The research interaction has a short range (and thus is very strong) centered at whichever ``lab''\footnote{Lab is a loose term, ``office'' might be more appropriate} each GS finds themselves pulled into. This interaction is well understood, and has the following form.

\begin{align}
    V_{\text{research}} &=s -  \ln(d)^{2}e^{-d}
\end{align}

Where $s$ is again the Cocky Stress Index, and $d$ is the distance from the lab \footnote{If you're worried about units here, you should be - how a GS experiences space and time is a current unsolved problem}. If $E_{gs} \leq V_{\text{research}}$ the grad student cannot stray far from the lab.  This interaction is so strong that most GS do not bother leaving the lab at all, since they do not want to spend the energy doing so. However there is a purely quantum effect called ``Grad-Student Tunneling", where GS are observed at coffee shops, art galleries, grocery stores, well outside the potential barrier from the research interaction. By solving the Schr\"{o}dinger equation, we can get a sense of this effect.

\begin{align}
    \psi &= A e^{-\sqrt{\frac{2m_{e}(V-E_{gs})}{\hbar^{2}}}d}
\end{align}

Where $A$ is an undetermined constant. If we know the specific boundary conditions of the excursion, the probability of barrier penetration can be found. 
\\

An exception to this strong research interaction is the ``FF'' (free food) interaction. It seems that during times of extremely low motivation, a GS may be able to freely leave the lab if there exists a FF effect. The FF effect is similar to the lab interaction, where GS are drawn towards particular locations, however the FF interaction is short lived. As the GS(s) reach the source of the FF, the potential drops quickly to zero, and the research interaction once again takes over and GS are spontaneously dropped back down to the lab state\footnote{With however, higher attitude, but also higher procrastination}. Some experiments have shown a relationship between the intensity of the FF interaction, stress index,  and the distance from lab.

\begin{align}
    V_{l\ell}/ |\text{FF}| &=\left(s + d\right) e^{-sd}
\end{align}

Where $V_{l\ell}$ is the potential that the GS will leave the lab, $s$ is the stress index, and $d$ is the distance from the lab.

\begin{figure}[h!]
    \centering
    \begin{tikzpicture}
        \begin{axis}[
        axis lines = left,
        xlabel = Distance from lab $d$,
        ylabel = {$V_{e\ell}$},
]
        \addplot[
        domain=0:10, 
        samples=150, 
        color=red,
                ]
                {(1+x)*exp(-x)};
        \addlegendentry{$s=1$}
        \addplot[
        domain=0:10, 
        samples=150, 
        color=blue,
                ]
                {(2+x)*exp(-2*x)};
        \addlegendentry{$s=2$}
\end{axis}
\end{tikzpicture}
    \caption{\label{fig:1}}
    \end{figure}

Figure \ref{fig:1} shows two GS stress levels, as you can see - the higher the $s$ index, the more potential a GS has to leave the lab, but they're not willing to travel as far (because the research potential well is so deep). Finally we can discuss the ``Food to Travel Ratio'' (FTR)

\begin{align}
    \text{FTR} &= \frac{|\text{FF}|}{d}
\end{align}

The FTR of $1$ is a normalized measure to a stress index $s=1$ indicating that the GS can only just break the research potential barrier to leave the lab. FTR greater than $1$ (Meaning the amount of FF (free food) is much larger than the distance from the lab) indicates that the GS may have an easier time leaving the lab.

\subsubsection*{GS/GS Interaction}

Until now we have been avoiding the tricky issue of GS/GS interactions, because of the phenomenon discussed previously, these interactions are difficult to observe. Qualitatively, GS do not interact with other GS very strongly. For instance, two unbound GS might not interact at all unless it is energetically (research-getically) favorable, and even then it is rare\footnote{More on this in statistical GS theory}.\\

However, it seems that within the lab, small groups of three to five GS may bind in something called a ``Poincare Research Group'', which exhibits special symmetries with respect to research translation and reflection (copying) that are beyond the scope of this paper. With this in mind, we may move onto the ever mysterious ``Advisor'' interaction.

\subsection{Advisor Interaction}
The advisor interaction (AI) is one of the most studied problems in all of Quantum Student Theory because of it's high influence on GS motivation, attitude, and decay. The AI does not occur on any nice periodic timescale, no does it have a centralized spatial dependence. In fact, it is increasingly difficult to observe the AI carrier directly, as they interact most strongly with the ``Post-Doc'' mentioned earlier, loosely with GSs, and almost not at all with UGs.
\\

The motivation for the AI is this, consider a lab of GS in which has high attitude, but low motivation and low productivity. Suddenly, upon observation the system flips - the lab as a whole has extremely high productivity, extremely low attitude, and wildly varying motivation - this is the ``Advisor Effect'.
\\

Qualitatively we may describe the AI as being carried by a theoretical A-particle, which as mentioned before interactions most strongly with PDs (post docs) and most weekly with UGs. We may draw the conclusion that the A-particle discriminates based on total progress, where similar total progress is energetically favorable. When the A-particle interacts with one or more GS, the GS are excited to a very high productivity state that will decay in both time, and distance from the A-particle. Furthermore, the A-particle tends to scatter from GS to GS without losing any energy whatsoever (which is why we might observe an entire lab being effected). 
\\

The A-particle interacts with UGs extremely rarely, and only when two criteria are met - the UG is about to decay into a GS, and the UG is interacting via the research interaction, or the lab potential. Because of UGs low productivity, the A-particle interacts with them more weakly, however their attitude state still drops to the zero-point level.
\\

As an additional mystery, A-particles have never been observed in labs. We \textit{know} the interaction takes place based on measurables from the lab system, but it seems that the A-particle is purely quantum in nature and is not observed at such motivations\footnote{As researchers usually have}. This leads to a second interpretation of the AI, which proposes that the AI is mediated by a A-field rather than a particle. The A-field permeates all space, but is particularly strong in labs. However, this proposal does not explain the irregularly periodic nature of the interaction. 
\\

Both hypotheses tend to fail when one considers the UG/GS interaction. Generally the UGs and GS only interact through the class interaction, or sometimes the research interaction. In the latter case, it seems after an AI to the GS, the GS will produce a similar effect on the UG, but to a lesser extent. This is called the UG/GS symmetry.

\subsection{Spooky Action At A Seminar}
It has been observed that GS and AI quantum can interact in another way called ``Seminar''(WS)\footnote{WS, because seminars tend to be periodic with a week between each occurrence}. When observed, a GS during the WS interaction will display both a high productivity, and a high procrastination - how does this happen? 
\\

During the WS interaction, the GS is still acted upon by the A-particle (or A-field), thus increasing its productivity. However, the GS exhibits a duality of procrastination during this time - this effect has been coined ``Surfing the internet'', where the GS is both productive during the WS interaction, and unproductive at the same time. In the end, the GS shows no new progress, but the AI has been fulfilled (and thus the A-particle drops back to it's low energy state). However, this interaction is spooky, because at any moment the A-particle may interact with the GS, and the GS will always display a state of productivity\footnote{independent of motivation, distance or time}.

\section{Statistical Student Theory}
Having touched on QST, we can now look into a brief overview of the statistics of non-interacting student quanta and the differences in how these quanta occupy states of discrete motivation\footnote{It should be noted that all further calculations assume productivity equilibrium}.
\\

Experiments using complicated equipment\footnote{Probably shooting UGs through magnets} have shown that all students exhibit a property called ``Sleep'' $S$. The ``Sleep'' is unrelated to the common notion of sleep, as student's don't actually show this behavior, rather, $S$ is related to the concept of angular sleep\footnote{In which a student falls asleep at a desk}.

\subsection{B-E statistics and the UG}
As luck would have it, UG's are perfectly described using Bose-Einstein statistics because of their integer $S$ values. Consider two UGs (whose names do not matter), which can be in two possible equi-motivation states $\ket{\alpha}$ and $\ket{\beta}$. Since UG are treated as identical, this system only has three different states.

\begin{align}
    \ket{\alpha}\ket{\alpha}, \quad \ket{\beta}\ket{\beta},\quad \frac{\sqrt{2}}{2} \bigg[ \ket{\alpha}\ket{\beta} + \ket{\beta}\ket{\alpha} \bigg]
\end{align}
And thus the probability any of these states will occur is equal (specifically about a third). This is yet another valid way to explain the CEF exhibited by UGs, as it is statistically probable that UGs will pile into the same state of motivation in equilibrium. 

\subsection{F-D Statistics and the GS}
Contrary to the UG statistics, GS have $S = 1/2$ or multiples thereof, and are thus described by Fermi-Dirac statistics. Consider the same motivation state scenario with two undistinguished GS. There is only one allowed antisymmetric state,

\begin{align}
    \frac{\sqrt{2}}{2} \bigg[ \ket{\alpha}\ket{\beta} + \ket{\beta}\ket{\alpha} \bigg]
\end{align}

with GS having a probability of $1$ of occupying this state. This state is often called the ``Single GS state'', or ``Singlet'', where it is motivationally favorable for two and only two GS to pair into a motivation state\footnote{and also the same spatial location}. This fits well with our observations of small numbers of GS coupled together.

\section{Conclusion}
We have traveled through many differing motivation and procrastination regimes to end our journey with an overview of the dynamics and interactions of the UG and GS. It has been discovered that the UG decays into GS, and both are subject to well defined interactions, and interact loosely with each other. We have also seen that quantum effects are observed, and new ``weird'' interactions like the AI are needed to explain such effects. 
\\

QTS still has a lot of progress to make, there are questions such as ``How did the first University come to be?'', ``What was before the University?'', ``Did the first university start with UG, GS, and A-particles, or was there some super-symmetric particle which decayed into UG?'' which still remain to be answered. Fields such as Condensed Education Theory, String-Along-Students theory, Quantum Studentology all aim to discover more of the secrets that lie beneat the University system to truly discover what it means to be a ``Ph.d''.
\end{document}